\documentclass[12pt]{article}
\usepackage{amssymb}

\usepackage{amsmath}
\usepackage{enumerate}
\usepackage{natbib}

\newcommand{\f}{\mbox{\textbf{f}}}
\newcommand{\g}{\mbox{\textbf{g}}}

\newcommand{\barE}{E^C} 

\newcommand{\I}{\mbox{I}}
\newcommand{\II}{\mbox{II}}
\newcommand{\III}{\mbox{III}}
\newcommand{\IV}{\mbox{IV}}
\newcommand{\U}{{\cal U}}

\begin{document}

\title{\textbf{On rereading Savage}}
\author{
Yudi Pawitan$^1$ and Youngjo Lee$^2$\\
$^1$Department of Medical Epidemiology and Biostatistics,\\
Karolinska Institutet, Stockholm 17177, Sweden (e-mail: yudi.pawitan@ki.se)\\
$^2$Department of Statistics, Seoul National University,\\
Gwanak-gu, Seoul 08826, South Korea (e-mail:
youngjo@snu.ac.kr).}

\maketitle
\subsection*{Abstract}
If we accept Savage's set of axioms, then all uncertainties must be
treated like ordinary probability. Savage espoused subjective
probability, allowing, for example, the probability of Donald
Trump's re-election. But Savage's probability also covers the
objective version, such as the probability of heads in a fair toss
of a coin. In other words, there is no distinction between objective
and subjective probability. Savage's system has great theoretical
implications; for example, prior probabilities can be elicited from
subjective preferences, and then get updated by objective evidence,
a learning step that forms the basis of Bayesian computations.
Non-Bayesians have generally refused to accept the subjective aspect
of probability or to allow priors in formal statistical modelling.
As demanded, for example, by the late Dennis Lindley, since Bayesian
probability is axiomatic, it is the non-Bayesians' duty to point out
which axioms are not acceptable to them. This is not a simple
request, since the Bayesian axioms are not commonly covered in our
professional training, even in the Bayesian statistics courses. So
our aim is to provide a readable exposition the Bayesian axioms from
a close rereading Savage's classic book.

\noindent \textbf{Keywords:} Axioms of probability; Bayesian;
subjective probability; uncertainty

\newpage

\section{Introduction}\label{sec:intro}
With a regular use of the inverse probability method, 19th-century
statistics was largely Bayesian. But Bayesian school of statistics
as a fully coherent statistical philosophy and methodology emerged
during 1950s, perhaps conveniently coinciding with the publication
of Savage's seminal book \emph{The Foundation of Statistics} in
1954. Even the adjective `Bayesian' was only used regularly in its
current meaning then after the appearance of many prominent
Bayesians that seemed to revolve around Savage (Fienberg, 2006).
Fisher (e.g., 1930),  following many 19th century mathematicians
including George Boole, John Venn and George Chrystal, criticized
the use of inverse probability method due to its arbitrary prior
probability for unknown parameters. Fisher had used the term
Bayesian pejoratively to refer to this method. Savage's work
established Bayesian statistics axiomatically, hence brought
legitimacy to prior probability and subjective probability in
general.

There is a virtual consensus regarding the use of probability for
statistical modeling, but we have yet to reach that for its
interpretation and philosophical aspects. Mathematically,
Kolmogorov's axiomatic foundation puts probability as the legitimate
child of the more mature measure theory. Kolmogorov's probability, with
its celebrated laws of large numbers, is naturally interpreted as
long-run frequency. In orthodox introductory teaching, where most
classic textbooks are written by frequentists, probability is said
to be meaningless for specific events such as Donald Trump's re-election in 2024. But people do bet on such specific
events, which can only mean that they do have a probability that is not of a long-term variety. The reasoning requires a probability that applies to specific events. Moreover, since different people have different beliefs and temperaments, they may have different probabilities for the same event. This in turn calls for subjective probability.

Savage's axiomatic development is a culmination of the subjective
theory of probability at the heart of Bayesian statistics, so it
seems appropriate to put our focus on him. In order not to get too
distracted, we shall not discuss Jeffreys (1939) or the development
of the so-called objective Bayesian statistics. Rather than
`subjective', Savage liked to use the term `personalistic,' but the latter is not in common use, so we shall use `subjective'
throughout. Our aim is to provide a brief historical background of the subjective probability, then to go through Savage's axioms in detail. We discuss two well-known paradoxes: Allais's and Ellsberg's. The purpose is not to use them in order to dismiss the axioms, but to illuminate the logical implications of some of them.

\section{Historical background}
There are two historical strands leading to Savage: (i) Ramsey's (1931) and de Finetti's (1937, RDF) definition of subjective probability as a betting quotient, and their fundamental theorem that the bet is coherent if and only if the probability satisfies finitely additive probability axioms. (ii) von Neumman and Morgenstern's (VNM, 1947) theorem in game theory that a person whose preferences follow certain axioms behave as if he is maximizing a utility function. In effect the RDF theorem assumes objective utility (good ol' money) and a coherent betting strategy to arrive
at subjective probability. This result is not sufficient to deal with the question of Bayesian prior probability. The VNM theorem assumes objective probability and axioms of rational preferences to arrive at subjective utility. Savage came up with the axioms of rational preferences that unify both subjective probability and subjective utility within the decision framework. This axiomatic
approach justifies the use of subjective probability, including
subjective priors, in Bayesian statistics.

Let's start with the work of Ramsey in Cambridge and de Finetti in Italy during 1920s. In the Cambridge circle, under the influence of
philosophers W.\ E.\ Johnson and Bertrand Russel, Keynes (1921)
proposed a logical theory of probability as a rational degree of
belief of propositions rather than just of events; `rational' is
meant to be objective, independent of subjective preferences. But
his idea, which still relied on the classical principle of
indifference/non-sufficient reason for equiprobable alternatives,
never made any impact in statistics or in mathematics. It is of
interest historically as it led Ramsey in 1926, who was only 22 at the time, to propose his alternative theory of probability as a subjective degree of belief. During 1920s de Finetti also conceived the same idea independently of Ramsey, though his book was only published in 1937. In their construction, a person's probability is revealed as a `coherent betting quotient', a value that avoids an external agent to run a Dutch Book -- a risk-free bet -- against him.

Setting a `coherent bet' is reminiscent to 'I-cut-you-choose' method of splitting a cake fairly between two greedy individuals. Even children will immediately see its fairness: John cuts the cake and Patrick chooses first. It is in John's self-interest to cut it as
equally as possible. To cut into unequal pieces is not coherent, as
he would knowingly guarantee himself to lose as Patrick will choose
the bigger piece.

How do we extend this trick to probability of an arbitrary event,
such as re-election of Donald Trump? Say John has a subjective
probability $0.25$ for such an event. It is not important that he
knows it himself, but only that he is prepared to act -- to bet --
on it. It is the action, speaking louder than any verbal
pronouncement,  that is supposed to reveal the probability. We set
up the following betting game. John (as player) has the chance to
win \$100 from Patrick (as bookie) if the event occurs, but he has
to pay Patrick a price to play the game. How much would he pay?
Paying more than \$25 would violate his sense of the probability of
the event, so he will not do that, but he would be happy to pay,
say, \$10. But that will of course not reveal his probability.

Here is the trick: Patrick \emph{has the right to reverse} their
roles as player and bookie depending on the amount John decides to
pay. If John decides to pay \$10, then (he worries) he runs the risk
that Patrick would reverse their roles and instead make him pay
$\$100$ in return for \$10 if the event happens. So, to neutralize
this risk, John should pay as large as possible, but as he would not
be willing to pay more than \$25, so he will arrive at \$25 as his
fair price. At this fair price John should be indifferent to the
role reversal. Choosing any other number runs the risk of a loss: if
his subjective probability is $p$ and he sets the price at $q$ per
winning dollar, he runs the risk of losing $|p-q|$.  So the
subjective probability $p$ might also be called the risk-neutral
(not risk-free) probability.

Several assumptions are implicit in this idealized betting game. One
is precision: in real life John the player will not have a precise
price in mind, so precision is a mathematical convenience.  At his
fair price, it is assumed that John would be indifferent between
being a player or a bookie. In practice he may have different
preferences, i.e. the chance of winning \$100 feels different from
the chance of losing \$100, even if they are balanced by the same
price or compensation. A real-life Patrick the bookie will have his
own probability of the event and will then set a price higher than
that. For example, he may accept even \$10 if that is higher than
his probability. So, the revelation of John's probability requires
that Patrick does not have his own price, while John must think
introspectively and reflexively that Patrick is thinking like
himself so that the risk of reversal occurs even when John is
willing to pay \$20. This condition is satisfied, for example, if
both presume that they have access to the same background
information.

Furthermore, it must be assumed that the amount of money involved is
small relative to their wealth, so that they can still think
linearly. In large amounts the utility of money becomes  non-linear,
so the reasoning must account for risk aversion. Finally, the game
might feel contrived with its potential role reversal and various
assumptions. Role reversal means you can be either a buyer or a
seller of a bet. This is actually what happens in online betting
exchanges, where you can choose to buy or sell depending on the
difference between your subjective probability and the current
`market price' of an event. More generally, it corresponds to buying
and shorting an asset in the financial market.


Under those same assumptions, this setup can be extended to an
arbitrary but finite number of complementary events, where John
specifies the price for each event according to his subjective
probabilities. Ramsey and de Finetti proved the fundamental theorem
that a betting strategy is coherent if and only if the probabilities
follow the axioms of finitely additive probability. Much has been
written about the finite additivity of subjective probability, as
opposed to the countable additivity of Kolmogorov's probability. In
fact, this is not a serious issue. Williamson (1999) provided a
proof of countably additive subjective probability under one extra
condition that only a finite amount of money is used, a condition
that is already tacitly assumed in order to avoid non-linear utility
effect. The assumption of fixed linear utility, however, renders the
RDF theorem insufficient as the basis of Bayesian statistics, where
a general loss function -- equal to minus utility -- would be
needed.

\section{Savage's axioms}
\subsection{The elements}
Savage conveniently listed his seven axioms as the `Postulates of a
Personalistic Theory of Decision' at the front end of his book. The
`formal subject matter' includes (i) $S$ the set of `states of the
world' $s_1,s_1',\ldots$; (ii) $F$ the set of `consequences'
$f_1,f_1', \ldots$, (iii) the set of `acts' $\f_1, \f_2,\ldots$.
Formally, an act $\f$ is a function from $S$ to $F$. We might write
$\f(s) = f(s)$ as a consequence at state $s$, but we have preserved
Savage's explicit preference for these notations. To facilitate
understanding -- and to tie up more easily with Ramsey, de Finetti,
and von Neumann and Morgenstern -- think of an `act' as a bet on the
state of the world, and the `consequence' is the payoff. (In general
the payoff can be expressed in utility scale, but for now there is
no need to do that yet.)

The `state of the world' is a description sufficient to compute the
consequences of each act. Perhaps useful to envision the usual
random experiment with $S$ as the sample space, but, unlike in von
Neumann and Morgenstern's lottery, don't assume that the random
generating mechanism is known. For example, imagine putting a bet in
a horse race: the bet and the payoff are obvious, and the sample space is the list of horses. The process of producing a winner is a
complex random process; the theory expresses a person's probability model with his betting behavior.

An abstract term `state space' -- rather than `sample space' -- is
used as the theory makes no distinction between the uncertainty
involving unknown fixed parameters, where no random experiment is
involved, and the uncertainty associated with truly random outcomes.
(In the case of fixed parameters, pedagogically, it is usually
presented as if the true parameters are sampled from a subjective
probability distribution, but technically no sampling is needed.) A
state is said to `obtain', or to `be true', if it is the true
parameter in the parameter space $S$, and can be interpreted as
`realized' if it is a randomly generated state in the sample space
$S$. For example, the winning horse is the `true state' among the
participating horses.  Subsets of $S$ are called events. An event is
said to `obtain' if it contains the true state; in standard
probability terms, we say the event `occurs.' We shall use the
latter term, because `obtain' is still not in common usage.

\begin{table}[h!]
    \centering
    \begin{tabular}{cccc}
 Bet  & Horse A & Horse B & Horse C \\
   \hline
  $\f_1$ & \$100 & \$0 & \$0 \\
  $\f_2$ & \$0 & \$100 & \$0 \\
  $\f_3$ &\$0 & \$25 & \$25
    \end{tabular}
    \caption{\label{tab:horses}A race of three horses A, B and C, and
        the corresponding payoffs associated with the bets $\f_1$,
         $\f_2$ and $\f_3$.
        }
\end{table}

Table~\ref{tab:horses} shows a simple example that captures the
elements of the theory. Choosing the bet $\f_1$ means you believe
Horse A would win, or more realistically, you believe Horse A has a
higher chance to win than the other horses. Similarly, choosing
$\f_2$ means you believe in Horse B; choosing $\f_3$ means you
believe either B or C would win. As in Ramsey's and de Finetti's
theory, a betting choice reveals a subjective probability. In one
key difference with standard probability model: a horse race may
already be completed. As will be clear from the axioms, the whole
theory applies unchanged to the bets made on the specific result of
a race, even a race that has concluded, of course as long as the
result remains unknown to the bettor. In contrast, the standard
probability theory does not apply to the result of a specific race.

The final element of the theory is a preference ordering between the
acts: `$\f_1 \le \f_2$' means `$\f_1$ is not preferred to $\f_2$,'
or more cleanly `$\f_2$ is preferred to $\f_1$; this is equivalent
to $\f_1<\f_2$ or $\f_1= \f_2$, meaning either $\f_2$ is strictly
preferred over $\f_1$ or they are equivalent. Now we are ready for
Savage's seven postulates.

\subsubsection*{Axiom 1}

\begin{quote}
 P1 (Weak ordering)\ The preference order is complete and
transitive, meaning that (i) every pair $\f_1$ and $\f_2$ can be
compared and ordered, (ii) for every $\f_1$, $\f_2$ and $\f_3$, if
$\f_1 \le \f_2$ and $\f_2 \le \f_3$, then $\f_1 \le \f_3$.
\end{quote}

This postulate is similar to von Neumann and Morgenstern's (1947)
Axioms 1 and 2 for subjective utility.  Transitivity is a crucial
element of any axiomatic system of rational behavior. A typical
concern is on the completeness: we can imagine situations where we
put little thoughts, hence have no explicit preferences, over some
acts. In statistics this is not likely to be a concern, and Savage
advised against considering a system that allows partial ordering.
As a simple consequence of the axiom, if $F$ is finite, then there
must be an optimal act, i.e. there is no other act strictly
preferred over it.

Savage already from the beginning pointed out two aspects of an
axiomatic system: empirical and normative. The former is as a
descriptive and predictive theory of people's behavior in decision
making: the axioms will be considered good or bad according to the
closeness of the predictions to actual behavior of people. The
normative aspect attempts to guide people's behavior, particularly
behavior that is not consistent with the theory. He brought analogy
to logic, where once we agree to certain propositions, then we
should follow their logical consequences. Axiomatic systems such as
Savage's or von Neumann and Morgenstern's carry their supposed
normative values from this iron law of logic. Savage's correction of
his own reaction of Allais's paradox below shows that he considered
his system normatively. So in this normative sense, the axioms are
axioms of rational behavior.

\subsubsection*{Axiom 2}
\begin{quote}
 P2 (Sure-thing principle)\ For every act $\f, \g$ and event $E$,
 $\f \le \g$ given $E$ or $\g\le \f$ given $E$.
\end{quote}

This axiom at the front-end of the book looks deceptively simple,
raising the question why it is called the sure-thing principle and
why it is controversial. The idea of conditional preference as used
in the axiom requires care, because different pairs of acts might
agree when $E$ occurs, but do not agree when $E$ does not occur.
Conditional preference is defined as follows:
\begin{quote}
 $\f \le \g$ given $E$, if and only if $\f' \le \g'$ for every  $\f'$
 and $\g'$ that agree with $\f$ and $\g$, respectively, on $E$, and
 with each other on $\barE$ and $\g' \le \f'$ either for all such
 pairs or for none.
\end{quote}

Now that is not so simple. Within the text, the postulate gets an
alternative longer version, not requiring a definition:
\begin{quote}
 P2 (Sure-thing principle)\ If $\f$, $\g$, and $\f'$, $\g'$ are such
 that:
 \begin{description}
 \item{1.} in $\barE$, $\f$ agrees with $\g$, and $\f'$ agrees with
 $\g'$,
\item{2.} in ${E}$, $\f$ agrees with $\f'$, and $\g$ agrees with
 $\g'$;
 \item{3.} $\f \le \g$;
  \end{description}
 then $\f' \le \g'$.
\end{quote}

Using a definition in one place and an axiom in another is an
interesting stylistic choice; in the latter the conditional
preference is an undefined primitive concept characterized in the
axiom. In case you think this longer version still looks
challenging, Savage illustrated it with an example (Savage, 1972,
page 21):
\begin{quote}
`A businessman contemplates buying a certain piece of property. He
considers the outcome of the next presidential election relevant to
the attractiveness of the purchase. So, to clarify the matter for
himself, he asks whether he would buy if he knew the Republican
candidate were going to win, and decides that he would do so.
Similarly, he considers whether he would buy if he knew the
Democratic candidate were going to win, and again finds that he
would do so. Seeing that he would buy in either event, he decides
that he should buy.'
\end{quote}
He then added a famous comment that `I know of no other extralogical
principle governing decisions that finds such ready acceptance,'
which is why he called the postulate `the sure-thing principle.'
Yet, mapping the example to the postulate is not obvious, because
the postulate involves four distinct acts, while the businessman
contemplates only two. In fact, the example is a more appropriate
illustration of the so-called `dominance principle,' which involves
consistent ordering of two acts across the states of the world.

\begin{table}[h!]
    \centering
    \begin{tabular}{cccc}
 Bet  & Horse A & Horse B & Horse C \\
   \hline
  $\f$ & \$100 & \$0 & \$0 \\
  $\g$ & \$0 & \$100 & \$0 \\
  $\f'$ &\$100 & \$0 & \$100\\
  $\g'$ & \$0 & \$100 & \$100
    \end{tabular}
    \caption{\label{tab:horses-P2}A race of three horses A, B and C, and
        the corresponding payoffs associated with four bets labeled
        according to the acts in postulate P2. The pair ($\f,\g$) is
        a choice between A vs B, while the pair ($\f',\g'$) is between
        not-B vs not-A. So the two pairs are logically distinct
        choices; P2 is a consistency requirement that they
        have the same preference ordering.}
\end{table}

Let's consider instead the example in Table~\ref{tab:horses-P2}
involving four bets in a horse race. The bet $\f'$ is a bet on
horses A and C, while $\g'$ is on B and C. Define the event
$E\equiv$ \{A,B\}. The pairs of bets $(\f,\g)$ and $(\f',\g')$ agree
on $E$, while on the complement event $\barE$ the bets agree within
each pair. In words, if $E$ does not occur, the two bets within each
pair are equivalent; while if $E$ occurs, the two pairs are
indistinguishable in their payoffs. Thus P2 specifies that if, given
$E$, you decide $\f \le\g$ then you must also decide $\f' \le\g'$,
and vice versa. This sounds reasonable: in the within-pair
comparisons of the bets, only horses A and B contribute preferential
information, C is irrelevant because it produces the same payoff
within each pair.

Versions of postulate P2 appear in other axiomatic systems. It is
closely related to Rubin's postulate or Milnor's column linearity
postulate (Luce and Raiffa, 1957), which states that adding a
constant amount to a column of payoffs -- associated with one state
-- should not change the ordering between acts. If the probabilities
of the events are known, as in a lotto, P2 is equivalent to von
Neumann and Morgenstern's (1947) fourth axiom, called the
independence axiom, stipulating that the decision maker compares
alternatives based on their distinct consequences, and ignores
aspects that are the same. The postulate also appears as Samuelson's
special independence assumption (Samuelson, 1952). In social choice
theory (e.g., Arrow, 1950), it is related to the axiom of
independence of irrelevant alternatives.

\subsubsection*{Axiom 3}
\begin{quote}
 P3 (State-independence)\ If $\f\equiv g$, $\f'\equiv g'$,
 and event $E$ is not null, then
 $\f \le \f'$ given $E$, if and only if $g\le g'$.
\end{quote}

First define the preference ordering of consequences $g$ and $g'$ in
terms of constant acts $\f\equiv g$ and $\f'\equiv g'$; think of
these acts as receiving gifts instead of betting. The axiom states
that the ordering of consequences remains the same under any event,
or that knowledge of the underlying event does not change the
preferential ordering of consequences. P3 is followed by a
significant theorem: assuming axioms P1 and P2, axiom P3 is
equivalent to the admissibility or the dominance principle, a key
principle in Abraham Wald's non-Bayesian statistical decision
theory. In brief, an act is not admissible if there is another act
that dominates it in terms of its consequences. The principle states
that we must reject non-admissible acts; e.g. we should reject bets
whose payoffs at each state are worse than those of another bet.

Leading to P3, Savage also defined a null event, which is an event
under which all acts are equivalent; for example, when all bets have
the same payoff, so the decision maker could ignore it from further
preferential considerations. According to this definition, a null
event is not an impossible event in the usual probabilistic sense,
but one that the decision maker does not care enough that it makes
no impact on his preferences. For example, for someone who does not
care to write his will and has no preference what happens to his
wealth after he dies, all the life-threatening events are null
events. It also means that for the states that matter, i.e. all the
non-null states, there must be at least one consequence that is
distinct from the others. This will appear in postulate P5.

Since in the limit an event can contain a single state, the axiom
can be called ordinal state independence of payoffs. Thus the
preferential value of \$100 payoff means the same regardless of the
underlying state that produces it. This seems appropriate for
statistical applications, but there are situations where it may not
be the case. In general the axiom leads to state-independent and
chance-neutral utility. In health economic or insurance
applications, a person's utility of money changes depending on his
underlying state of health. Chance neutrality means, the value one
puts to, say, \$1000 is the same regardless whether it comes from
the regular salary or from winning a lotto. Recently Stef\'ansson
and Bradley (2015) discussed this aspect in rationality and utility
theory.

\subsubsection*{Axiom 4}
\begin{quote}
 P4 (Consequence independence)\ If consequences $f,f',g,g'$;
  events $A, B$; and acts $\f_A, \f_B,  \g_A, \g_B$ are such that:

 \begin{tabular}{llll}
 {1.} & $f'<f$, & $ g'<g$\\
 {2a.} & $f_A(s) = f$, &  $g_A(s)=g$ & for $s \in A$\\
       & $f_A(s) = f'$, &  $g_A(s)=g'$ & for $s \notin {A}$\\
{2b.} & $f_B(s) = f$, &  $g_B(s)=g$ & for $s \in B$\\
       & $f_B(s) = f'$, &  $g_B(s)=g'$ & for $s \notin {B}$\\
3. & & $\f_A \le \f_B$;
\end{tabular}

then $\g_A \le \g_B$.
\end{quote}

The act $\f_A$ can be interpreted as a bet on event $A$, as it
offers a strictly better payoff if $A$ occurs; similarly for the
other three acts. So $\f_A \le \f_B$, i.e. preferring $B$ over $A$
even though they have the same payoffs, must mean you believe $B$ is
more likely than $A$. The axiom states that the choice of events to
bet on is independent of the size of the payoff. In the horse race
example (Table~\ref{tab:horses}), the choice of Horse B over A
should remain the same if the payoff is changed to $\$200$ or any
other value. Axioms 3 and 4 are the key pillars that support a great
weight: the separation between subjective probability and utility,
or between belief and value, which are normally tangled in ordinary
preferences. This means (i) utility of payoffs can be defined
independently of the underlying state, including its probability,
and (ii) the subjective probability of an event is meaningful
independently of the value of the event. The former can be
interpreted as chance neutrality: when you win say \$1M in a
lottery, you should feel the same regardless whether the winning
probability is close to impossible $10^{-9}$ or just small
$10^{-6}$. Recently Stef\'ansson and Bradley (2015) propose the idea
of chance itself having a utility.

\subsubsection*{Axiom 5}
\begin{quote}
 P5 (Non-triviality)\ There is at least one pair of consequences $f,
 f'$ such that $f'<f$.
\end{quote}

This must the case under one state of the world. In view of the
discussion of null events in Axiom 3 above, this axiom is needed to
avoid a trivial null state space. In other words, there is at least
one event, containing one state, that you are willing to bet on.

\subsection{Qualitative probability and Axiom 6}

Axioms 1-5 form a natural set, as they are sufficient to construct a
\emph{qualitative subjective probability.} It is similar to the
logical probability concept that appeared in Keynes's proposal,
which he considered necessary when there is no sufficient knowledge
to set up quantitative probability. In this framework we can say,
for example, that `event $A$ is less probable than $B$' without
specifying their numerical probabilities. In his 1941 paper
`\emph{Heuristic Reasoning and the Theory of Probability}' and
delightful 1954 book `\emph{Mathematics and Plausible Reasoning},'
the mathematician G.\ Polya also described a coherent setup of
qualitative logical probability as a tool for characterizing the
experimental stage of mathematical proofs.

Defining 0 as null event, $S$ as the state space, a relational
operator $\preceq$ is a qualitative probability if, for all events
$B, C$ and $D$,
\begin{description}
\item{1.} $\preceq$ is complete and transitive,
\item{2.} $B\preceq C$ if and only if $B\cup D\preceq C\cup D$,
provided $B\cap D= C\cap D=0$,
\item{3.} $0\preceq B$, $0\preceq C$.
\end{description}

It is, however, more convenient to express the relation in
probability notation. We first define a finitely additive
probability measure as a function of subsets $B$ of $S$, such that
\begin{description}
\item{1.} $P(B)\ge 0$ for every $B$.
\item{2.} If $B\cap C=0$, $P(B\cup C) = P(B) + P(C)$.
\item{3.} P(S)=1.
\end{description}
The probability measure $P$ and the qualitative probability
$\preceq$ are said to agree if for every $B$ and $C$, $P(B)\le P(C)$
is equivalent to $B\preceq C$. Unfortunately Axioms 1-5 are not
sufficient to guarantee the existence of a probability measure that
agrees with $\preceq$. Savage proved a series of theorems leading to
postulate P6', which provides the guarantee.

\begin{quote}
P6' If $B\prec C$, there exists a partition $\cup_i E_i$ of $S$,
such that $B\cup E_i \prec C$ for any $i$.
\end{quote}

The key idea here is that $S$ is rich enough so that it can be
divided into a fine partition, such that any piece in the partition
is too small to change the strict ordering of $B$ and $C$. Savege
viewed this postulate as a precursor to postulate P6 needed for a
full-blown quantitative subjective probability.

\begin{quote}
P6 (Small-event continuity)\ If $\f < \g$, and $h$ is any
consequence; then there exists a partition $\cup_i E_i$ of $S$ such
that, if $\f$ is so modified to take value $h$ on $E_i$ for any $i$,
other values being undisturbed, then the modified $\f <\g$. The same
ordering is also preserved when $\g$ is modified instead.
\end{quote}

The postulate requires $S$ to be rich enough, at least infinitely
countable, to allow a fine partition, and each act is a smooth
function over $S$ that a modification on a small event does not
upset the preference of the act. Typically this is interpreted to
mean there is no consequence either infinitely better or infinitely
worse than the others that its inclusion in a modification of an act
upsets its ordering. Another consequence is that the probability is
\emph{non-atomic} in the sense that for any $\rho$, $ 0\le \rho\le
1$, and event $E$, there is $B\subseteq E$, such that $P(B)=\rho
P(E).$

Allais's paradox below shows that a small-event discontinuity can
occur, even with bounded consequences, if the small event turns an
act into a sure bet. The continuity axiom is closely related to von
Neumann and Morgenstern's third axiom, sometimes called Archimedian
property for no obvious reasons. It is a continuity requirement in
the set of acts -- lotteries -- in that no act is infinitely more or
infinitely less preferable than any other act. Further technical
work is needed to deal with finite state spaces as in the horse race
example. Wakker (1989), for example, proposed a model where the set
of consequences is a connected separable topological space.

Axioms 1--6 are sufficient to construct quantitative/numerical
subjective probability, which conveniently reduces the calculation
of preferences between acts in terms of arithmetic comparisons of
numbers. However, the acts must be of special form, i.e. they are
bets on events. So what we have is numerical probability of any
event. At this point the theory is difficult to navigate, since
there is no explicit theorem stating exactly what has been achieved
by the time we reach Axiom 6. Comparisons of arbitrary acts with
arbitrary consequences will have to wait until the introduction of
utility and Axiom 7.

\subsubsection*{Finite additivity}

Similarly to Ramsey-de Finetti's coherent betting strategy, Savage's
axioms 1-6 are necessary and sufficient to construct finitely
additive probability. This is in contrast to Kolmogorov's countably
additive probability, which is the basis of most results in modern
probability theory. However, agreeing with de Finetti, Savage knew
`of no argument leading to the requirement of countable
additivity... therefore seems better not to assume countable
additivity outright as a postulate ... .' Kolmogorov  (1933) himself
argued that the countable additivity axiom was essential
theoretically, but `it is almost impossible to elucidate its
empirical meaning.'

Assuming countably additive probability, we cannot have a uniform
distribution over, say, positive integers. De Finetti (1970, vol 1,
p-122) considered that counter-intuitive. However, his reasoning was
not clear. Countable additivity `forces me to choose some finite
subset of them ... to which I attribute a total probability of at
least 99\%....' Savage (1972, page 43) also wrote `many of us have a
strong intuitive tendency to regard as natural probability problems
about the necessarily only finitely additive uniform probability
densities on the integers, on the line and on the plane.' Under
finite additivity, we can of course have a uniform distribution on
an arbitrarily large collection of integers, but the collection must
be finite, so it is not clear how finite additivity solves the
problem.

The loss of countable additivity is a serious loss. For example,
Schervish et al (1984) showed that we lose conglomerability. Under
countable additivity, for any event $E$ and partition $\cup_i B_i$
of $S$, $P(E)$ must be in the interval between $\inf_i P(E|B_i)$ and
$\sup_i P(E|B_i)$. This is intuitively compelling, but it is not
true if we only have additivity. As with the Ramsey-deFinetti's
subjective probability, there is a known remedy to Savage's version
also: Villegas (1964) added a monotone continuity condition to make
the probability countably additive.

\subsection{Conditional probability}

Conditional probability can be constructed starting with conditional
preference as described in Axiom 2. If $\preceq$ is a qualitative
probability relation, then for events $B$ and $C$, and non-null
event $B$, we can define the conditional relation $B\preceq C$ given
$D$ to mean $B\cap D \preceq C\cap D$. It can then be shown that the
conditional relation is also a qualitative probability. If $\preceq$
is numerically represented by (`almost agrees with') the probability
measure $P(\cdot)$, then the conditional relation is represented by
$$
P(B|D) = \frac{P(B\cap D)}{P(D)}.
$$
This is of course the key formula that allows update of subjective
probability, the crucial step of learning from experience, which
forms the basis of Bayesian inference. Savage then moved
imperceptibly from the qualitative to a full quantitative
conditional probability; presumably all follows from Axioms 1--6.

Then came (i) a familiar introduction of the Bayesian method with
updating of prior probability $P(B_i)$ of event $B_i$, using data
$x=(x_1,\ldots,x_n)$, to produce posterior probability
$$
P(B_i|x)\propto P(x|B_i) P(B_i).
$$
And (ii) an exposition of exchangeability based on de Finetti's
example on the inference of the success probability $p$ from a
sequence of Bernoulli trials. The classical objective view is that
$p$ is a fixed unknown parameter. For de Finetti, `objective' is
simply an agreement of subjective opinions. Different people may
start with different opinion -- hence different priors -- but their
posterior probability will eventually agree, so there is no need for
the objective view. In both expositions there is an emphasis on
consistency as $n$ tends to infinity, and there is no discussion on
how to get the prior distribution.

\subsection{Utility, Axiom 7 and the theorem}

Up to this point, acts that are bets on events can be compared by
numerical probabilities. But we cannot compare acts with arbitrary
consequences, which is of course necessary for general statistical
applications. To do that we need the concept of utility $\U$, such
that
\begin{equation}
\f \le \g\ \mbox{if and only if}\ \U(\f) \le \U(\g). \label{eq:util}
\end{equation}
In this case we say that the preference ordering agrees with the
utility. Given a probability measure $P(\cdot)$ on $S$, the utility
$\U(\f)$ is the expected value
$$
\U(\f) \equiv \int_S U(f(s)) dP(s),
$$
where $U(\cdot)$ is the utility function applied to consequences
$f(s)$. The utility concept allows a convenient arithmetic
representation of acts and their preference ordering. If we consider
the axioms normatively as axioms of rational decisions, then the
utility function is a calculator of rationality, so it has special
role in any discussion of rationality and economic behavior.

Savage's utility theory is largely influenced by von Neumann and
Morgenstern's (1947, VNM) utility. In the final foundational chapter
(Chapter 5), Savage discussed the utility concept at great length,
providing the historical background from the time of Daniel
Bernoulli in the 18th century, and defending it from the then
orthodox economic view that had dismissed any meaningful value of
cardinal -- as opposed to ordinal -- utility. He was siding strongly
with von Neumann and Morgenstern, and perhaps even going beyond them
by interpreting the theory normatively as a guide of rational
behavior.

As we have seen above, three of the six axioms are in fact closely
related to VNM's three axioms to establish the existence of utility
as the basis of preferential ordering of lotteries. By first
defining a gamble as an act with a finite set of consequences,
Savage first proved that Axioms 1--6 are sufficient to guarantee the
existence of a utility function that agrees with the preferences
between gambles. So Axiom 7 is introduced to allow theoretically
infinite number of consequences:
\begin{quote}
P7\ (Strong dominance) If $\f \le g(s)$ for every $s$ in $B$, then
$\f\le \g$ given $B$.
\end{quote}
This means, if every consequence of $\g$ is preferable to $\f$ as a
whole, then $\g$ must be preferable to $\f$.  Savage considered this
as a stronger version of the sure-thing principle. P7 is clearly a
stronger version of the dominance principle, which only compares the
consequences $f(s) \le g(s)$ at each state $s$. As we mentioned
above, Savage proved that, assuming P1 and P2, postulate P3 is
equivalent to the dominance principle. This means that P7 makes P3
redundant in the whole set of axioms (cf.\ Hartmann, 2020).

\subsubsection*{The theorem}

The utility concept completes Savage's axiomatic development, and
the overall theorem is stated as follows: The preference ordering of
acts satisfies the postulates P1--P7 if and only if there exists a
unique finitely additive non-atomic probability measure $P$ on S and
a real-value bounded utility function $U(\cdot)$ on $F$, such that
the preference ordering agrees with expected utility in the sense of
(\ref{eq:util}).

This means someone who acts rationally, in the sense that his
preferences satisfy the axioms, behaves \emph{as if} he has a
probability measure and a utility function, and he maximizes the
expected utility. In principle he does not have to be even aware of
his subjective probability and utility function. However, in
practice it is much easier to first assume explicitly the
probability and the utility function, then work out the implied
preferences. For example, in Bayesian statistics, one typically starts
with a prior distribution and a minus squared-error loss as utility.
According of the theorem, maximizing the expected utility
corresponds to choosing an optimal act. By staying within the
convenient utility framework, in other words `by following the
expected utility theory,' your preferences are guaranteed
`rational.' Any decision that violates the expected utility can be
shown to violate at least one of the axioms.

Savage's theorem can be seen as an amalgamation of Ramsey-de
Finetti's subjective probability and von Neumann-Morgenstern's
utility. Strictly speaking, when the probability is objective -- as
in a lotto -- the utility theory is covered in von
Neumann-Morgenstern's framework, otherwise it is in Savage's.

\section{Paradoxes}

We now discuss two well-known paradoxes, not in order to dismiss the
axioms, but to illuminate the logical content and implications of
some of them.

\subsection{Allais's paradox}

The French economist and nobel laureate M.\ Allais (1953) described
two betting situations where people tend to prefer alternatives that
contradict the expected utility theory, hence violating at least one
of the axioms. It is an important paradox, because among the
violators, when Allais first presented it in a 1952 meeting on the
economic theory of risk, were future Nobel prize winners in
economics (Paul Samuelson, Milton Friedman and Kenneth Arrow).
Savage also happened to there; his reaction will be presented below.

\begin{description}
\item{Situation 1.} Choose between these two bets (it's important
to use these large amounts of money, because utility is
magnitude-dependent, though one can modify \$2,500,000 with smaller
amounts such as \$1,000,000 and preserve the paradox):

\begin{tabular}{ll}
  I. & Win \$500,000 with probability 1,\\
  II. & Win \$2,500,000 with probability 0.1,\\
        & or \$500,000 with probability 0.89\\
        & or nothing with probability 0.01.
  \end{tabular}
\item{Situation 2.} Choose between these two bets:

  \begin{tabular}{ll}
  III. & Win \$ 500,000 with probability 0.11,\\
         & or nothing with probability 0.89.\\
  IV. & Win \$2,500,000 with probability 0.1,\\
        & or nothing with probability 0.9.
  \end{tabular}
\end{description}

Assuming a linear utility function, the expected utility of bet II is
$$
\U(\II) = 0.1\times 2,500,000 + 0.89\times 500,000 = 695,000,
$$
which is greater than $\U(\I)=500,000$, so you should prefer II. But in reality, most people would be happy with the smaller but sure amount of money in option I. The small potential of getting nothing in bet II is not compensated by the another small potential of a bigger jackpot. So people tend to prefer I over II. This is  a well-known risk aversion, but, wait, it is not yet the paradox in question.

How about between III and IV? A small reduction in the probability of winning \$500,000 in III is rewarded by a large increase in the jackpot in IV. As expected, most people would indeed prefer IV over III. \textit{Now} we have a paradox: that choice turns out to violate the expected utility theory. Kahneman and Tversky (1979) conducted many
surveys of people's preferences in bets similar to Allais's and
confirmed the offending preferences.

The usual explanation of the paradox is nothing to do with
non-linear utility nor with risk aversion. We can first put the four bets in terms of their expected utilities for a general non-linear utility
function $U(\cdot)$ (for convenience, let's set $U(0)\equiv 0$, and
drop 000s from the numbers):
\begin{eqnarray*}
\U(\I)&=& U(500)\\
\U(\II) &=& 0.1U(2,500) + 0.89U(500)\\
\U(\III) &=& 0.11 U(500)\\
\U(\IV) &=& 0.1 U(2,500)
\end{eqnarray*}
Preferring I$>$II implies $U(500)>0.1 U(2,500)+0.89U(500)$, or
$0.11U(500)
> 0.1 U(2,500)$. But the last preference ordering means you should
prefer III over IV, which contradicts people's common preference of
IV over III.

The common preference is usually cited as a violation of  the
sure-thing principle. Imagine 100 lottery tickets numbered 1 to 100,
where you pick one ticket, and a winning number is chosen randomly.
The payoffs are given in Table~\ref{tab:allais}. Then the four bets
in the table are exactly equivalent as above, and we can see clearly
the set-up of the sure-thing principle. Under the principle, the
last column (tickets 12-100) represents the irrelevant event, so the
preference I over II must be consistent with III over IV.
\begin{table}[h!]
\centering
\begin{tabular}{lrrr}
Bet & 1 & 2--11 & 12--100\\
\hline
\I & 500 & 500 & 500 \\
\II & 0 & 2500 & 500\\
\III & 500 & 500 & 0\\
\IV & 0 & 2500 & 0\\
\end{tabular}
\caption{\label{tab:allais} Allais's four bets are equivalent to the
bets in this table, where the columns indicate ticket numbers and
the entries are the payoffs. You pick one number between 1 and 100,
and a winning number is selected randomly. So, the probability of
winning is 0.01, 0.10, and 0.89 for the three columns,
respectively.}
\end{table}

Savage initially chose like most people do  (i.e., preferring I over
II, and IV over III), but after realizing that he violated the
sure-thing principle, he changed his preference to III $>$ IV. He
wrote (Savage, 1972, page 103) `in reversing my preference between
Gambles 3 and 4 I have corrected an error.' But, why did he reverse
III and IV, not I and II? The sure-thing principle only states that
the two pairs of bets must be consistent, but does not say which one
is the right choice. In any case, Savage's correction highlights the
normative value of the axioms as a guide for rational decision, with
power to correct our subjective preferences. It raises an
interesting question whenever there is a conflict: which have the
primacy? Do we modify our preferences or modify our axioms? We
discuss this further in the Discussion Section.

\begin{table}[h!]
 \centering
\begin{tabular}{lrrrr}
Bet & 1 & 2--11 & 12--100\\
\hline
\I & 500 & 500 &  500\\
$\I_0$ & 0 & 500  & 500\\
\II & 0 & 2500 & 500\\
$\III$ & 500 & 500 &  0\\
$\III_0$ & 0 & 500 &  0\\
\IV & 0 & 2500 & 0
\end{tabular}
\caption{\label{tab:allais2} Modified bets $\I_0$ and $\III_0$ in
Allais's paradox, following Table~\ref{tab:allais}, where the payoff
of some tickets are set to zero. For clarity, the original bets are
also listed. }
\end{table}

The explanation of Allais's paradox as a violation of the sure-thing
principle is well-known, but we can also argue that the paradox
violates the small-event continuity axiom (Axiom 6). Viewed in this
perspective, the paradox occurs because of a conflict between the
axiom and risk aversion. Let's modify the payoff of some bets on
events with small probability, i.e. on ticket 1 with probability
0.01, to get $\I_0$ and $\III_0$. These modifications are shown in
Table~\ref{tab:allais2}. To be clear, the reasoning steps are
itemized as follows:
\begin{itemize}
\item By the strong-dominance principle (Axiom 7), we must have II $> \I_0$
and IV $>\III_0$.
\item III is a small-event modification of III$_0$ using a comparable-sized
payoff, so by Axiom 6, we should have IV$>$III. This is the commonly
observed preference, so in fact it is justified by Axioms 6 and 7.
\item Now, the modification I$_0\rightarrow\I$ is exactly the same as
III$_0\rightarrow\III$, so again by Axiom 6, we should have II$>$I.
However, the small-event modification from I$_0$ to I generates a
sure gain, so for most people the ordering II $> \I_0$ gets reversed
to I $>$II, thus violating Axiom 6.
\end{itemize}
In this analysis, the preference IV$>$III is seen not as an error,
but a choice that agrees -- at least in spirit if not formally --
with Axioms 6 and 7. So, as we remarked above, there is actually no
rational basis for Savage to reverse his initial preference of
IV$>$III. The reason he reversed it was because he first preferred
I$>$II, which was justified by risk aversion. But risk aversion is
an empirical phenomenon that is not implied by any of the axioms and
plays no role in Savage's theory. So, normatively, reversing the
risk-averse preference I$>$II is perhaps a more consistent choice.
But you would sacrifice risk aversion in order to follow Axiom 6.

Instead of using Axiom 6, we could also invoke the sure-thing
principle to justify II$>\I$, but here we want to highlight the
violation of Axiom 6. We could also say: Axiom~6 and the sure-thing
principle together contradict risk aversion. Or, risk aversion and
the sure-thing principle together contradict Axiom~6. So, to the
extent that the small event in Axiom~6 is an approximation of real
events with small probability, the axiom is in conflict with risk
aversion. It is an important conflict, because risk aversion is
closely connected to the universally accepted law of diminishing
returns. In view of the discussion following Axiom 6, the reversal
of preference occurs even when there is no event with infinitely
better or infinitely worse consequence. The reversal -- marking a
discontinuity -- occurs when even a small-event modification creates
certainty, as people behave differently when dealing with sure events.

Such a violation is not a rare or exotic phenomenon;  as described
by Kahneman and Tversky (1979) it is commonly seen, for example, in
gambling and insurance decisions. They took the paradox seriously
and developed an alternative axiomatic system called the prospect
theory, in part to allow for discontinuity when the probability is
near 0 or near 1. Instead of maximizing expected utility, in
prospect theory a decision maker maximizes the score
\begin{equation}
V\equiv \sum_i \pi(p_i) U(f_i),\label{eq:V}
\end{equation}
where $U(f_i)$ is a function of the payoff $f_i$ (they called it `value function,' but in principle it works like a utility
function), $p_i$ is the probability of state $i$, and $\pi(\cdot)$
is a decision weight function with $\pi(0)\equiv 0$ and
$\pi(1)\equiv 1$. In the expected utility theory $\pi(p_i)=p_i$, but in general $\pi(p)\ne p$.  An important feature of the theory is that the decision weight function has a discontinuity property:
$\pi(0^+)>0$ and $\pi(1^-)<1$, which implies a sub-certainty
property $\pi(p) + \pi(1-p) <1$ when $0<p<1$. Furthermore, the prospect score (\ref{eq:V}) can be re-written as
$$
V = \sum_i p_i \{ \pi(p_i) U(f_i)/p_i\} \equiv \sum_i p_i
U^*(f_i,p_i),
$$
where $U^*(f_i,p_i)\equiv \pi(p_i) U(f_i)/p_i$ becomes a
chance-dependent or state-dependent utility function, thus violating
Axiom 3.

Prospect theory is considered one of the  cornerstones of behavior
economics; Kahneman was awarded Nobel prize in economics in 2002 for
this work. In particular, the theory provides better explanations of
distinct human reactions to gains and losses, and of behavior near
impossibility or certainty. Allais's paradox can be seen as being
due to special behavior associated with sure gain in Bet I. The
scores are
\begin{eqnarray*}
V(\I)&=& U(500)\\
V(\II) &=& \pi(0.1)U(2,500) + \pi(0.89)U(500)\\
V(\III) &=&\pi(0.11) U(500),\\
V(\IV) &=& \pi(0.1) U(2,500).
\end{eqnarray*}
The preferences I$>$II and IV$>$III imply:
\begin{eqnarray*}
\{1-\pi(0.89)\}U(500)&>& \pi(0.1) U(2,500) \\
\pi(0.1) U(2,500) &>& \pi(0.11) U(500),
\end{eqnarray*}
giving $\pi(0.89)+\pi(0.11)<1$, which is the sub-certainty property
anticipated (and allowed) in the prospect theory.

We close this section with another explanation of the paradox, highlighting an unstated assumption in the setup of the sure-thing principle. Let's assume that you have to make \emph{two simultaneous bets}: (I vs II) and (III vs IV) \emph{based on a single draw of the lottery} in Table~\ref{tab:allais}. Preferring I$>$II and IV$>$III implies (I+IV)$>$(II+III), which is irrational because they have exactly the same payoffs. The other preference (I+III) vs (II+IV) depends on one's utility of money, so there is no immediate inconsistency issue. This means that for the sure-thing principle to hold, the bets are presumed to be made simultaneously based on a single realization of the random outcome. Violators of the principle -- which include many brilliant economists -- are likely thinking of two independent bets, based on two random draws of the lottery. Had the dependence of the single draw of the lottery been made explicitly, a rational person is not likely to violate the principle.

\begin{table}[h!]
\centering
\begin{tabular}{lrrr}
Bet & 1 & 2--11 & 12--100\\
\hline
\I+\IV & 500 & 3000 & 500\\
\II+\III & 500 & 3000 & 500\\
\hline
\I+\III & 1000 & 1000 & 500 \\
\II+\IV & 0 & 5000 & 500\\
\end{tabular}
\caption{\label{tab:allais3} Allais's paradox explained in terms of two simultaneous bets on the outcome of a single draw of a lottery. Preferring I$>$II and IV$>$III implies (I+IV)$>$(II+III), which is irrational because they have exactly the same payoffs.}
\end{table}

\subsection{Ellsberg's paradox}

Ellsberg (1961) set up a paradox that suggests that people treat
objective and subjective uncertainty differently. Briefly, in an
urn, there are 90 coloured balls: 30 are red and the other 60 are an
unknown mixture of black and yellow. You pick one ball from the urn
and are given the option of (I) getting \$100 if the ball is red, or
(II) getting \$100 if the ball is black. Which option would you
prefer? While the payoff table is similar to the bets in
Table~\ref{tab:horses-P2}, but the logical content is distinct
because of the single information on black and yellow balls here.

\begin{table}[h!]
    \centering
    \begin{tabular}{cccc}
   & Red & Black & Yellow \\
   \hline
  I & 100 & 0 & 0 \\
  II & 0 & 100 & 0 \\
    \end{tabular}
    \caption{The first situation from Ellsberg (1961):
    an urn contains 30 red balls and 60 black balls
    and yellow balls together, but with unknown proportion.
    You pick one ball from the urn; the table shows the
    payoffs based on the color of the ball. Do you prefer option I or II?}
    \label{tab:ellsberg_1}
\end{table}

Now consider the second scenario: Again, you pick one ball from the
urn and are given the option of (III) getting \$100 if the ball is
red or yellow; or (IV) getting \$100 if the ball is black or yellow.
Which option would you prefer now?

\begin{table}[h!]
    \centering
    \begin{tabular}{cccc}
   & Red & Black & Yellow \\
   \hline
  III & 100 & 0 & 100 \\
  IV & 0 & 100 & 100 \\
    \end{tabular}
    \caption{The second situation from Ellsberg (1961):
    similar setup as Table~1, but different payoffs.
    Do you prefer option III or IV?}
    \label{tab:ellsberg_2}
\end{table}

If you are like most people (e.g., Camerer, 1992), you would prefer
I over II, and IV over III.  It shows that people tend to prefer the
objective probability over the subjective one; behavioral economists
call this tendency \textit{ambiguity aversion}. This is clearly in
violation of the sure-thing principle, which requires that
preferring I over II implies preferring III over IV, and vice versa.
According to the principle, the presence of yellow balls should not
influence the preference between red and black balls.

In this example, the probability of picking a black or yellow ball
is between 0 to 2/3, but this probability is subjective. Savage's
theory implies that prior probability can be determined by
self-interrogation of subjective preferences. He supported de
Finetti that binomial probability can be interpreted in terms of
subjective probability alone. Let assume, as Bayes and Laplace did,
the probability of picking a black ball follows the uniform
distribution on 0 to 1. Then, Laplace's law of succession shows that
the probability of $k$ black balls is 1/61 for all $k=1,\ldots, 60$.
This distribution satisfies de Finetti's exchangeability, and was
discussed thoroughly in Section 3.7 of Savage's book. The expected
value of the number of black balls is 30, and if we perform the
experiment repeatedly the marginal chance of a black ball being
selected is 1/3.

Thus, if we treat the subjective probability objectively, then I and
II are equivalent; similarly III and IV. However, in a specific bet,
the number of black balls could be 15 or 45, or any number between 0
to 60. While the number of red balls is predetermined as 30, the
number of black balls is not determined but only has an expectation
30. Thus, the number of black balls has another layer of
uncertainty, an ambiguity. Because of this ambiguity people prefer I
over II. A Bayesian may say that he believes in his prior, so that I
and II are equivalent, but if so he is ambiguity-neutral and ignores
the extra uncertainty in a specific situation.

This paradox also highlights the descriptive and normative aspects
of an axiomatic system. If the system is only descriptive, then we
simply say that it does not predict well in this case. But if it is normative, as basic logic or arithmetic, then the violators must re-consider and correct their choices. Ellsberg (1961) reported various reactions among some
well-known economists and statisticians. Some did not violate the
principle (G.\ Debreu, R.\ Schlaifer, P.\ Samuelson). Some violated
the principle `cheerfully' (J.\ Marschak, N.\ Dalkey); others `sadly
and persistently, having looked into their hearts, found conflicts
with the axioms and decided to satisfy their preferences and let the
axioms satisfy themselves. Still others (H.\ Raiffa) tend,
intuitively, to violate the axioms but feel guilty about it ...'
Some who previously felt `first-order commitment' to the principle
were `surprised and dismayed to find that they wished, in these
situations, to violate the Sure-Thing Principle.' This special group
`seems to deserve respectful consideration' because  it included
none other than L.\ J.\ Savage himself.

As with the Allais's paradox, there is a setup where ambiguity  aversion is clearly
irrational. It requires one to play \emph{two simultaneous bets} I vs
II, followed by III vs IV \emph{based on a single draw from the balls.} First let's put a price on each bet: since I and IV have objective probabilities, assuming linear
utility, the fair prices are \$33 and \$67, respectively. If you have ambiguity aversion, you prefer I over II, and IV over III. This means you set lower for II (say \$30) and III (say \$65) relative to I and IV. If so, then I can run a Dutch Book against you: I will buy \emph{both bets} II and III from you for a total of
\$95, and with that I am guaranteed to win \$100. Now we can explain Ellsberg's paradox as follows: The sure-thing principle implicitly presumes the preferred acts to be acted together by one abstract decision maker based on the a single realization of the random outcome. But, unless stated explicitly, a real person
considers them one at a time as independent decisions based on independent draws from the balls. Explicitly requesting a rational and intelligent person to make two simultaneous bets based on a single draw will remove the aversion in this case. However, if the person is making independent single bets based on independent draws, it is not obvious whether ambiguity aversion violates the sure-thing principle.

\section{Discussions}

Savage dedicated a whole chapter to address criticisms of the
(personalistic) subjective probability as well as his own criticisms
of other views of probability. First, it is perhaps useful to go
through a number of statements that capture his views (page 46, 56,
57). We can recognize here his strong influence on modern
Bayesianism; see e.g. Lindley (2000).
\begin{quote}
...any mathematical problem concerning personal probability is
necessarily a problem concerning probability measures -- the study
of which is currently called by mathematicians mathematical
probability -- and conversely.

...the concept of personal probability introduced and illustrated in
the preceding chapter is... the only probability concept essential
to science and other activities that call upon probability.

...the role of mathematical theory of probability  is to enable the
person using it to detect inconsistencies in his own real or
envisaged behavior.
\end{quote}

\subsubsection*{Ambiguous probability}

Ellsberg's paradox (1961) highlights an example where people react
differently to different types of uncertainty, roughly as `sure' and
`unsure' or ambiguous probability. Savage dismissed the notion of
second-order probability to deal with such different levels of
uncertainty as in `the probability that $B$ is more probable than
$C$ is greater than the probability that $F$ is more probable than
$G$.'  The concept of ambiguous belief or `imprecise probability'
was later taken up, for example, by Schmeidler (1989), Gilboa and
Schmeidler (1993) and Binmore (2009)

Savage also did not support a model where the probability of $B$ is
a random variable $b$ with respect to the second probability. He was
concerned with `endless hierarchy' that would be difficult to
interpret, but the hierarchical models of this type have become popular
in recent years. In fact, it is now common to think of Bayesian
models as hierarchical, but Savage did not consider subjective
probability as part of a hierarchical model.

\subsubsection*{Inexact magnitude}

As with de Finetti's coherence argument, the postulates imply that
one can determine with high accuracy his subjective probability that
Donald Trump will be re-elected. In reality we can only expect some
rough magnitudes. Savage responded that the subjective theory, as a
normative theory, is `a code of consistency for the person applying
it, not a system of predictions...' De Finetti's (1931, page 204)
response to the same problem is also relevant here:  `...to apply
mathematics, you must act as though the measured magnitudes have
precise values. This fiction is very fruitful... To go, with the
help of mathematics, from approximate premises to approximate
conclusions, I must go by way to an exact algorithm...'

\subsubsection*{Other views of probability}

Savage considered the subjective probability as lying not in
between, but beside, the logical and objective views, because these
latter two are meant to be free of individual preferences. For him,
the strength of his axiomatic system is that it deals explicitly
with the problem of decision under uncertainty. He saw a weakness in
other subjective views, such as Koopman's (1940), that did not
explicitly deal with the problem of individual decisions. De
Finetti's coherent betting approach was criticized because it needs
to assume the utility of money is linear (or the bet is small). In
his later life de Finetti (1964) moved in the direction of Savage by
putting probability and utility together within a decision theory.

For the objective view -- i.e., frequentist -- probability is
primary, decision secondary. Furthermore, probability is only given
to very special events, e.g. repeatable ones such as coin tosses,
but not to very specific event such as the unification of Korea. For
Savage, the objective view is `charged with circularity.' It relies
on the existence of processes that can be represented by infinitely
repeatable events, but the degree of approximation is determined by
the same theory of probability. Savage also rejected the need for a
dualistic view of probability in inference, which allows both
objective and subjective ones.

\subsubsection*{Objectivity in science}

No one will stop you from betting your own money, but how can we
justify the subjective beliefs in science? We may define objectivity
as agreement between reasonable minds, excluding the possibility of
differing individual preferences. Savage put forward some arguments
why there is no reason to exclude subjective probability as part of
scientific reasoning. One argument is, by consistency, any two
differing opinions will be brought closer by sufficiently large
evidence. The objective view presumes a common universally accepted
opinion as the goal of science, but in reality there are `pairs of
opinions, neither of which can be called extreme ... which cannot be
expected to brought into close agreement after the observational
program' (Savage, 1972, page 68).

Ryder (1981) pointed out that an external agent can run a Dutch Book
against two individuals that hold differing subjective probabilities
of an event, i.e., make money from them regardless of what happens.
Say John has a subjective probability 0.25 for re-election of Donald Trump, while Patrick has 0.15. Alice comes along to bet
\$100 against John, and $-\$100$ against Patrick (meaning she plays
the bookie against John, but lets Patrick be the bookie; or, even more clearly, she buys from Patrick and sells to John). If Trump is re-elected, Alice's gain is
$$
g_1 = 0.25*100 - 100 -0.15*100 + 100 = 10,
$$
while, if he is not, she gains
$$
g_2= 0.25*100 - 0.15*100 = 10.
$$
The only way to avoid the Dutch Book is for John and Patrick to have
the same probability. (From the financial-market perspective, the two players behave like two mini-markets that offer different prices, which provide an arbitrage opportunity -- i.e. risk-free investment -- to an external agent.) Note that both John and Patrick are individually coherent, and if there is no communication between them, they are none the wiser about the Dutch Book run against them. But if they are closely related individuals with regular contacts, and especially with a joint economy, we could say they are incoherent,
because they have allowed a Dutch Book against them.

If it is not too farfetched of an analogy, a collection of
scientists in a scientific area is meant to be a closely communicating group
with a common interest. So if the individual scientists have
differing subjective probabilities of relevant events, the group can
be said to be incoherent. This might explain the reluctance of
nonBayesians or science in general to allow subjective preferences
in the \emph{formal assessment} of evidence.

\section{Conclusions}
An axiomatic system carries the strength of logic: once we agree
with a set of propositions, then we must agree with their
implications. It is therefore important that we understand Savage's
axioms in details. Savage clearly viewed Bayesian statistics not
simply about attaching a prior distribution to a statistical
problem, but a holistic framework for making rational decisions
under uncertainty. The framework is strong enough to carry a
normative force.

Axioms cannot be proved or disproved mathematically, but they can be checked empirically. For example, the fifth axiom in Euclid's
geometry -- the so-called parallel postulate -- can be corroborated
if the total sum of angles of the triangle in the space is 180
degrees. In a universe where the angles do not sum to 180 degrees
its geometry cannot be Euclidian. On the earth surface, large
triangles will have angles adding up to more than 180 degree. Thus,
the axioms can be checked via experience or observations and we may
encounter situations that violate the axioms. Shafer (1986) previously revisited Savage primarily to object the normative aspect of the theory, in light of accumulating empirical evidence that people violate the postulates.

For statistical applications, curiously Savage did not discuss the
problem of how to choose a prior distribution, which was already
considered by many writers from the 19th century as a weakness in
the inverse probability method. Much of this problem seems to be
addressed currently by the so-called objective methods, such as
Jeffreys's non-informative prior, but it is a substantial topic in
itself and way beyond the scope of this paper.

\section*{Acknowledgements}
This research was partly supported by the Brain Research Program
through the National Research Foundation of Korea (NRF) funded by
the Ministry of Science, ICT \& Future Planning (2014M3C7A1062896)
and the Korea government (MSIP) (No. 2011-0030810, No.
2019R1A2C1002408), and by the Korea-Sweden Research Cooperation
grant from the Swedish Foundation for International Cooperation in
Research and Higher Education (STINT).

\section*{\protect\normalsize References}

\begin{description}

\item Binmore, K. (2009). \emph{Rational Decisions.} Princeton: Princeton
University Press.


\item de Finetti, B. (1974). \emph{Theory of Probability} (Vol. I). London: Wiley.

\item Ellsberg, D. (1961b). Risk, ambiguity, and the Savage axioms.
\emph{Quarterly Journal of Economics}, \textbf{75}, 643--669.


\item  {\normalsize {Fisher R.A.}\ (1930). Inverse probability. \emph{Proceedings of the Cambridge Philosophical Society} \textbf{26}, 528--535. }


\item  {\normalsize {Fisher R.A.}\ (1958). The nature of probability. \emph{%
Centennial Review} \textbf{2}, 261--274. }

\item Gilboa, I., and Schmeidler, D. (1993) Updating ambiguous beliefs. \emph{Journal of Economic Theory}, \textbf{59}, 33--49.

\item Hartmann, L. (2020). Savage's P3 is redundant. Econometrica, \textbf{88}, 203?-205.

\item Kolmogorov, A.N. (1933) \emph{Grundbegriffe der
Wahrscheinlichkeitrechnung. }English translation by N . Morrison, \emph{Foundations of the Theory of Probability.} New York: Chelsea, 1956.

\item Koopman, B.O. (1940). The bases of probability. Bulletin of the American Mathematical
Society, \textbf{46}, 763--774.

\item Lindley, D.V.\ (2006). The philosophy of statistics. \emph{The Statistician}, \textbf{49}, 293--319.

\item Ramsey, F. (1931). \emph{Truth and Probability.} In F. Ramsey (Ed.), Foundations of mathematics and other logical essays. New York: Harcourt.

\item Ryder, J. M. (1981). Consequences of a simple extension of the Dutch Book argument. \emph{British Journal for the Philosophy of Science} 32, 164--167.

\item Savage, L. (1972). \emph{The Foundations of Statistics.} 2nd Edition. New York: Wiley.

\item Schervish, M., Seidenfeld, T. and Kadane, J. (1984). The extent of non-conglomerability of finitely additive
probabilities. \emph{Zeitschrift f\"{u}r Wahrscheinlichkeitstheorie
und Verwandte Gebiete}, \textbf{66}, 205-226.

\item Schmeidler, D. (1989). Subjective probability and expected
utility without additivity. \emph{Econometrica}, \textbf{57},
571--585.

\item Shafer, G. (1986). Savage Revisited. Statistical Science, \textbf{1},  463-485.

\item{Stef\'ansson, H.0., and Bradley, R.} (2015). How valuable are chances?.
\emph{Philosophy of Science}, \textbf{82}, 602-625.

\item Villegas, C. (1964). On Qualitative Probability $\sigma$-Algebras. \emph{Annals of Mathematical
Statistics,} \textbf{35}, 1787--1796.

\item Von Neumann, J., and Morgenstern, O. (1947).\emph{ The Theory of Games and Economic Behavior.} 2nd edition. Princeton: Princeton University
Press.

\end{description}

\end{document}